\documentclass[final,1p,times]{elsarticle} 
\usepackage{graphicx} 
\usepackage{amssymb} 
\usepackage{amsthm} 
\usepackage{lineno} 
\newcommand{\la}{\langle}
\newcommand{\ra}{\rangle}

\newcommand{\beq}{\begin{eqnarray}}
\newcommand{\eeq}{\end{eqnarray}}
\newcommand{\bfk}{\mbox{{\boldmath $k$}}}

\journal{Nuclear Physics A} 
\begin{document} 

\begin{frontmatter} 


\title{Critical Opalescence around the QCD Critical Point and Second-order Relativistic
 Hydrodynamic Equations Compatible with  Boltzmann Equation }

\author{Teiji Kunihiro$^{(a)}$, Yuki Minami$^{(a)}$ and Kyosuke Tsumura$^{(b)}$}

\address[a]{Department of Physics, Kyoto University, Kyoto 606-8502, Japan}
\address[b]{Analysis Technology Center, Fujifilm Corporation, Kanagawa 250-0193, Japan}

\begin{abstract} 
The dynamical density fluctuations around the QCD critical point (CP)
are analyzed using relativistic dissipative fluid dynamics, and 
we show that the sound mode around the QCD CP is strongly attenuated whereas
the thermal fluctuation  stands out there. We speculate that if possible
 suppression or disappearance
of a Mach cone, which seems to be created by the partonic jets at RHIC, 
is observed as the incident energy 
of the heavy-ion collisions is decreased, it can be a signal of the existence of the QCD CP.
We have presented the Israel-Stewart type fluid dynamic equations
that are derived rigorously  on the basis of the (dynamical) renormalization group method
in the second part of the talk, which
we omit here because of a lack of space.
\end{abstract} 

\end{frontmatter} 



\section{Introduction}\label{}

A unique feature of the QCD phase diagram
 is the existence of a critical point.
At the QCD CP, the first order phase transition terminates and turns to 
a second order phase transition.  Around a critical point of a second order transition,
 we can expect  large fluctuations of various quantities,
 and more importantly  there should exist a soft mode associated 
to the CP.
The QCD CP belongs to the same universality class as the liquid-gas phase 
transition point, and, hence, 
the density fluctuating mode in the space-like region 
 is a softening mode at the CP:
The would-be soft mode of the chiral transition, the $\sigma$ mode, is coupled to the 
density fluctuation\cite{Kunihiro:1991qu} and becomes a 
slaving mode of the density variable\cite{fujii};
 see \cite{Ohnishi:2005br} for 
another argument on the fate of the $\sigma$ mode around the CP.

The density fluctuation depends on the transport as well as thermodynamic
 quantities that show an anomalous behavior around the critical point. 
In particular, we should note that
 the density-temperature coupling which was not explicitly
taken into account can be important.
 In fact,  the dynamical density 
fluctuations are analyzed in the non-relativistic case with use of the Navier-Stokes equation,
which shows that the Rayleigh peak due to the thermal fluctuation would overwhelm the
Brillouin peak due to the sound modes\cite{reichl}.

We apply for the first time relativistic fluid dynamic equations to 
analyze the spectral properties of density fluctuations, and 
examine possible critical phenomena. 
We shall show that even the so called first-order 
relativistic fluid dynamic equations have generically no problem
to describe fluid dynamical phenomena with long wave lengths contrary to
 naive expectation.
In this report\cite{minami09},
 we shall show that the
genuine and remaining soft mode at the QCD CP is not
a sound mode but the diffusive thermal mode that is coupled to the sound mode,
and that the possible divergent behavior of the viscosities might not
be observed through the density fluctuations
because the sound modes are attenuated around the CP and would eventually
 almost die out at the CP.

\section{Relativistic fluid dynamic equations for a viscous system}

The fluid dynamic equations are 
the balance equations for energy-momentum and particle number,   
$\partial_\mu T^{\mu \nu}=0$,\,$\partial_\mu N^\mu =0$,
where $T^{\mu \nu}$ is the energy-momentum tensor
and  $N^\mu$  the particle current, respectively.
They are expressed as
$T^{\mu \nu}=(\epsilon+P)u^{\mu}u^{\nu}-Pg^{\mu\nu}+\tau^{\mu\nu}$
and  
$N^\mu = n u^\mu+\nu^\mu$, \,
where $\epsilon$ is the energy density, $P$  the pressure, $u^\mu$ the flow velocity, 
and $n$ the particle density, the dissipative part of the energy-momentum
tensor and the particle current are denoted by $\tau^{\mu \nu}$ and $\nu^\mu$,
respectively.

The so called first order equations such as Landau\cite{landau} and Eckart\cite{eckart}
equations are parabolic and formally violates the causality,
and are hence called acausal.
The causality problem is circumvented in the Israel-Stewart equation\cite{is},
which is a second-order equation with relaxation times incorporated.
One should, however, note
that the problem of the causality is only encountered
when one tries to describe phenomena 
with small wave lengths beyond the valid region of the fluid dynamics:
The phenomena which the fluid dynamics should describe
are  slowly varying ones with the wave lengths
much larger than the mean free path.
Indeed, the results for fluid dynamical modes
with long wave lengths are qualitatively the same 
irrespective whether the second-order or first-order equations are
used or not\cite{minami09}.
As for the instability seen in the Eckart equation\cite{hiscock}, a new first-order
equation in the particle frame constructed by Tsumura, Kunihiro and Ohnishi (TKO)
\cite{tko} has no such a pathological behavior.
We employ Landau\cite{landau}, Eckart\cite{eckart}, Israel-Stewart(I-S)\cite{is}
 and TKO equation.

\section{Spectral function of the dynamical density fluctuation}

By linearizing the fluid dynamic equation around the equilibrium,
 we can obtain the
spectral function of the density fluctuation.
The calculational procedure 
is an extension of the non-relativistic case described in the text book 
\cite{reichl}.

The spectral function derived from the Landau equation is found to be 
\begin{eqnarray}
S_{n n}(\bfk ,\omega ) 
   &=& \la (\delta n(\bfk ,t=0))^2\ra [\;(1-\frac{1}{\gamma})
   \frac{2\Gamma_{\rm R} k^{2}}{\omega^{2}+\Gamma_{\rm R}^{2}k^{4}}
   \nonumber \\
   &+& \frac{1}{\gamma}
   \{\frac{\Gamma_{\rm B} k^{2}}{(\omega -c_{s}k)^{2}+\Gamma_{\rm B}^{2}k^{4}}
   +\frac{\Gamma_{\rm B} k^{2}}{(\omega +c_{s}k)^{2}+\Gamma_{\rm B}^{2}k^{4}}\} \;].
   \label{eq:landau}
\end{eqnarray}
Here, the first factor  represent the static spectral function, which would
show a divergent behavior in the forward angle ($\bfk =0$) at the CP;
this is known as the critical opalescence.
The first term in the square bracket represents the thermal mode called Rayleigh mode,
whereas the second and the third the sound mode or Brillouin mode.

The Eckart equation in the particle frame does not give a sensible
result for the dynamical density fluctuation, in accord with its
pathological property\cite{hiscock}.
It is noteworthy that newly proposed equation, the TKO equation\cite{tko},
 in the particle frame
gives a sensible result even thou it is a first-order equation.
We have also applied the Israel-Stewart equation\cite{is} in the particle
frame to obtain the 
spectral function for the dynamical density fluctuation.
The result is the same as that of Landau equation; this tells us that
the modified part to circumvent the causality problem 
does not affect the dynamics in the proper fluid dynamic regime.

\section{Critical behavior of the dynamical density fluctuations}

We examine the critical behavior of the spectral function of the density fluctuations
around the QCD CP.
We introduce
the static critical exponents 
$\tilde{\gamma}$ and $\tilde{\alpha}$ which are defined as follows 
$\tilde{c}_n = c_0 t^{-\tilde{\alpha}}$,\,$K_T = K_0 t^{-\tilde{\gamma}}$,
where $t=\vert (T - T_c) / T_c \vert$ is a reduced temperature,
$c_0$ and $K_0$ are constants and $K_T=(1/n_0)(\partial n/\partial P)_T$ is the isothermal compressibility.
We also denote the exponent of the thermal conductivity by 
$a_{\kappa}$, i.e.,
$\kappa =\kappa_0 t^{-a_{\kappa}}$,
where $\kappa_0$ is a constant.
It is known that $a_{\kappa} \sim 0.6$ around the liquid-gas phase transition point.

\begin{figure}[htb]
\begin{tabular}{cc}
\begin{minipage}{0.5\hsize}
  \begin{center}
   \includegraphics[width=45mm, angle=270]{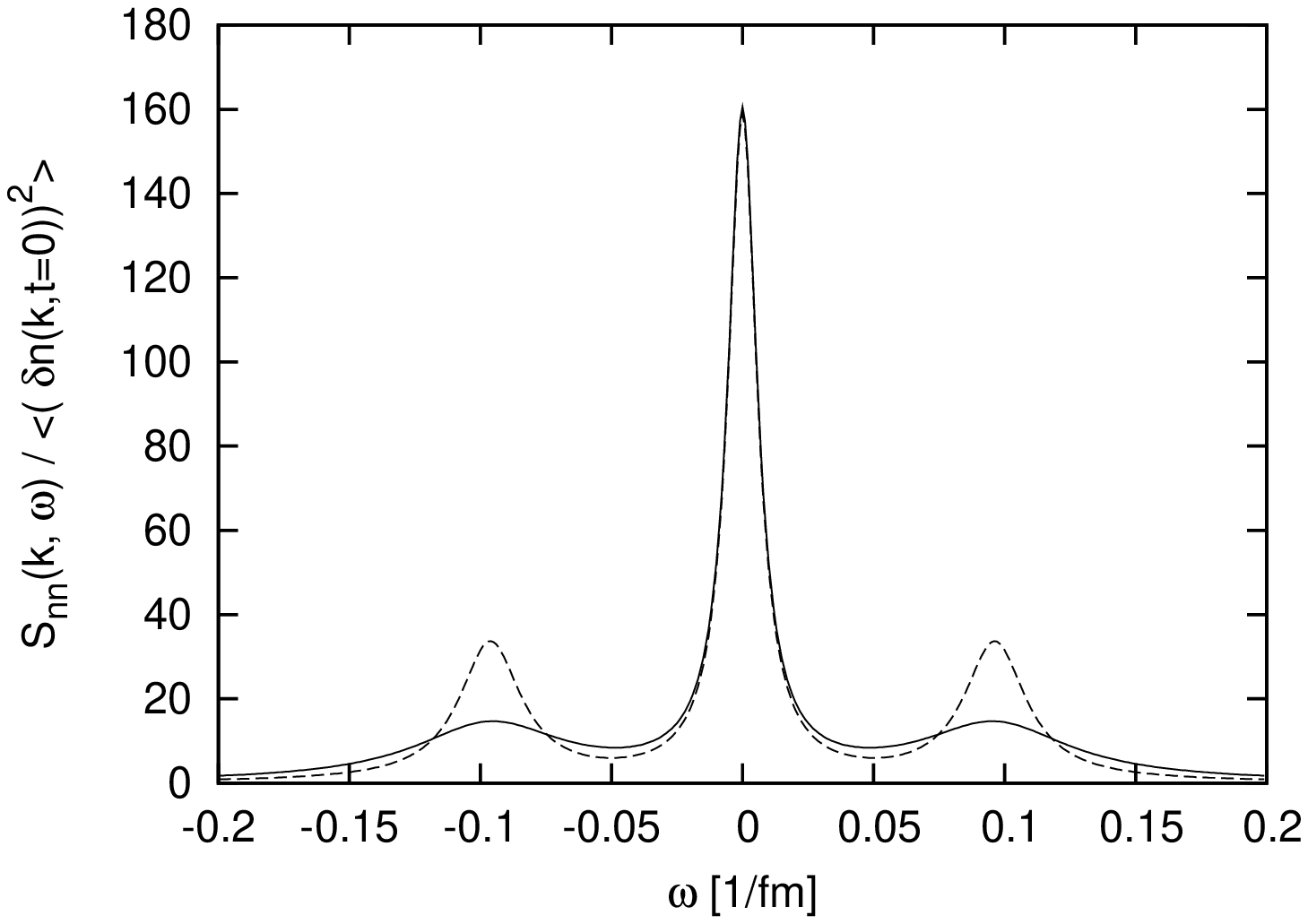}
  \end{center}
\end{minipage}
\begin{minipage}{0.5\hsize}
 \begin{center}
   \includegraphics[width=45mm, angle=270]{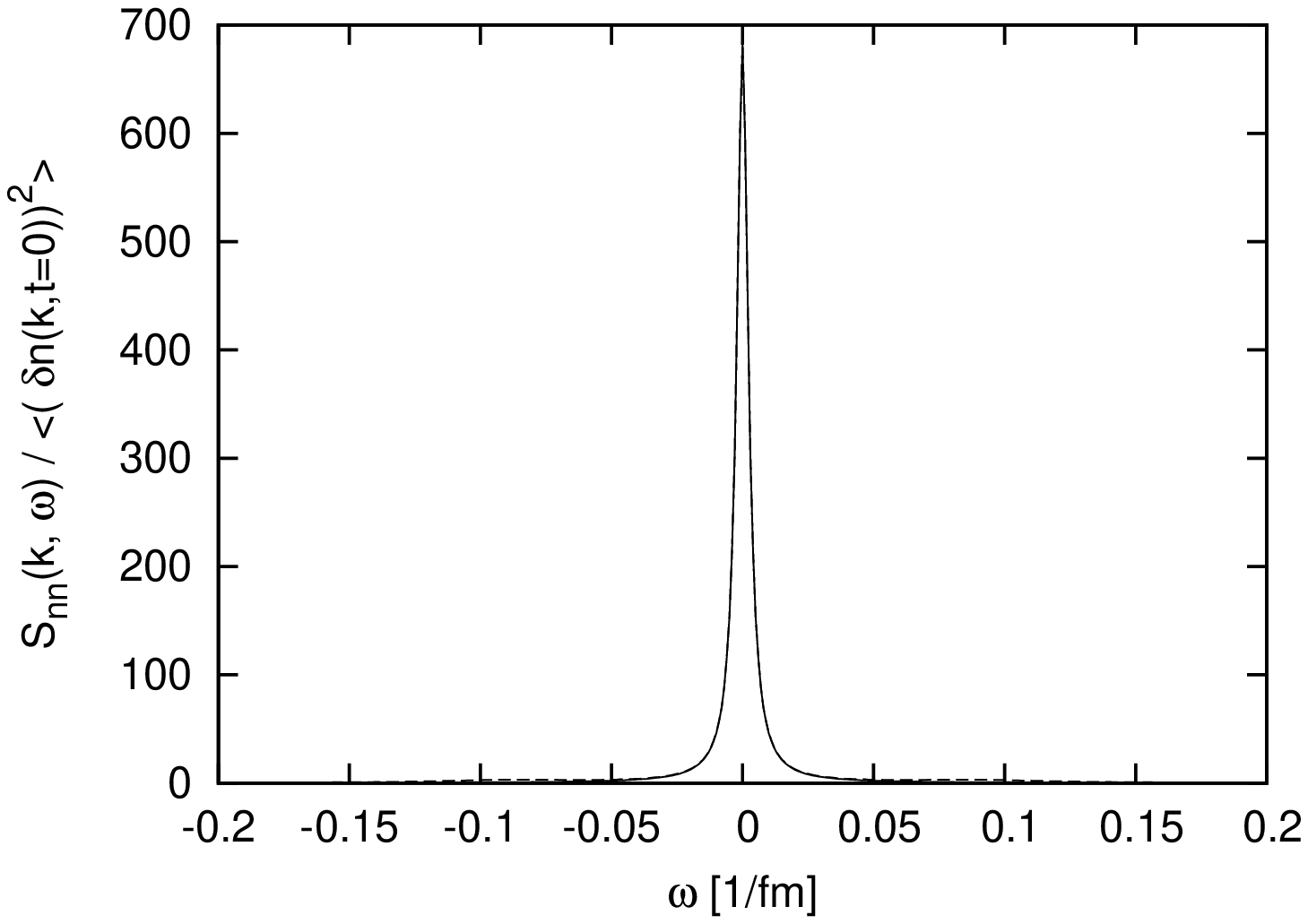}
\end{center}
\end{minipage}
\end{tabular}
\caption{The spectral function at $t\equiv (T-T_c)/T_C=0.5$\, (left panel)
 and at $t=0.1$ \, (right panel) for $k=0.1$[1/fm].
                      The solid line represents the Landau/Israel-Stewart case, while
           the dashed line  the TKO case. 
           The strength of the Brillouin peaks due to the machanical sound mode
 becomes smaller as $T$ approaches
$T_c$  due to the singularity of
 the ratio of specific heats;
the Brillouin peaks  eventually die out as seen from the right panel, where
the difference between the Landau and TKO cases is not seen anymore.
Note that the scale of the vertical line in the right panle is much bigger than that of
the left panel.}
  \label{fig:t5-1}
\end{figure}

Unfortunately or fortunately,
these singular behaviors of the width of the Brillouin peaks  around the
 QCD CP may not be observed.
The strengths of the Rayleigh and the Brillouin peaks
are given in terms of $\gamma$ as seen from eq.(\ref{eq:landau}),
 the ratio of the specific heats, which
 behaves like 
$\gamma = \tilde{c}_p / \tilde{c}_n \sim t^{-\tilde{\gamma}+\tilde{\alpha}}
\rightarrow \infty$,
in the critical region. 
Then the strength of the Brillouin peaks is attenuated 
and only the Rayleigh peak stands out in the critical region,
as shown in Fig.~\ref{fig:t5-1}.

Let $\xi=\xi_0t^{-\nu}$ be the correlation length which diverges
as the critical point is approached. 
If we write the wave length of the sound mode by $\lambda_s$,
the fluid dynamic regime is expressed as
$\xi << \lambda_s$,
with which condition the sound mode can develop.
However, in the vicinity of the critical point, the correlation length
$\xi$ becomes very large and eventually becomes infinity, so
the above inequality can not be satisfied, and the sound mode
can not be developed in the vicinity of the critical point.

From this argument, we can speculate about the
fate of  the possible Mach cone formation \cite{Torrieri:2009mv} 
by the particle passing through 
the medium with a speed larger than the sound velocity $c_s$.
Such a Mach-cone like particle correlations 
are observed in the RHIC experiment\cite{star}. 
Then the disappearance or suppression of
the Mach cone 
according to the lowering of the incident energy by RHIC
would be a signal of the existence 
of the QCD critical point provided that the incident energy is large enough to
make parton jets\cite{minami09}.

\section{Concluding remarks}

In this report, the density fluctuations is analyzed using the
relativistic fluid dynamic equations\cite{minami09}.
We have suggested that a suppression or disappearance of 
the Mach cone formation with lowering the incident energy at RHIC
can be a signal of the detection of the QCD CP.
Although we have presented the Israel-Stewart type fluid dynamic equations
that are derived rigorously  on the basis of the (dynamical) renormalization group method,
we omit them here because of a lack of space.
For the details, we refer to the submitted paper\cite{Tsumura:2009vm}, where it is shown that
the transport coefficients have no frame dependence while the relaxation times are generically
frame-dependent in the derived equations.

\section*{Acknowledgments} 
This work was partially supported by a
Grant-in-Aid for Scientific Research by the Ministry of Education,
Culture, Sports, Science and Technology (MEXT) of Japan (No.
20540265),
 and by the
Grant-in-Aid for the global COE program `` The Next Generation of
Physics, Spun from Universality and Emergence '' from MEXT.

\end{document}